\documentclass[submission,copyright,creativecommons]{eptcs}
\providecommand{\event}{TARK 2023} 
\usepackage{amssymb}
\usepackage{iftex}
\usepackage{times}
\usepackage{soul}
\usepackage{url}
\usepackage[utf8]{inputenc}
\usepackage[small]{caption}
\usepackage{graphicx}
\usepackage{amsmath}
\usepackage{amsthm}
\usepackage{booktabs}
\usepackage{algorithm}
\usepackage{algorithmic}
\newcommand{\LP}{\mathcal{L}}
\newcommand{\bs}[1]{{\bf #1}}
\newcommand{\MLP}{{\mathcal{M}_\LP}}
\newcommand{\mypostnoname}[2]{\noindent {\bf (#1)}\hspace{2mm}{#2}\hfill{\hspace*{0,1cm}}\\ }
\newcommand{\mypost}[3]{\noindent {\bf (#1)}\hspace{2mm}{#2}\hfill{ (#3)}\\ }

\newcommand{\bsrev}{{\star}}
\newcommand{\norm}[1]{\left\lVert#1\right\rVert}

\newtheorem{theorem}{Theorem}
\newtheorem{definition}{Definition}
\newtheorem{observation}{Observation}

\title{System of Spheres-based Two Level Credibility-limited Revisions}
\author{Marco Garapa
\institute{Universidade da Madeira}
\institute{CIMA - Centro de Investiga\c c\~ao\\ em Matem\'atica e Aplica\c c\~oes}
\email{mgarapa@staff.uma.pt}
\and
Eduardo Ferm\'e
\institute{Universidade da Madeira}
\institute{NOVA Laboratory for Computer Science\\ and Informatics (NOVA LINCS)}
\email{eduardo.ferme@staff.uma.pt}
\and
Maur\'icio D. L. Reis
\institute{Universidade da Madeira}
\institute{CIMA - Centro de Investiga\c c\~ao\\ em Matem\'atica e Aplica\c c\~oes}
\email{m\_reis@staff.uma.pt}}

\def\titlerunning{System of Spheres-based Two Level Credibility-limited Revisions}
\def\authorrunning{Marco Garapa, Eduardo Ferm\'e \& Maur\'icio D.L. Reis}
\begin{document}
\maketitle

\begin{abstract}
 {\em Two level credibility-limited revision} is a non-prioritized revision operation. When revising by a two level credibility-limited revision, two levels of credibility and one level of incredibility are considered. When revising by a sentence at the highest level of credibility, the operator behaves as a standard revision, if the sentence is at the second level of credibility, then the outcome of the revision process coincides with a standard contraction by the negation of that sentence. If the sentence is not credible, then the original belief set remains unchanged. In this paper, we propose a construction for two level credibility-limited revision operators based on Grove's systems of spheres and present an axiomatic characterization for these operators.
\end{abstract}

\section{Introduction}

{\em Belief Change} (also called {\em Belief Revision}) is an area that studies the dynamics of belief. One of the main goals underlying this area is to model how a rational agent updates her set of beliefs when confronted with new information. The main model of belief change is the AGM model \cite{AGM85}.  In that model, each {\em belief} of an agent is represented by a sentence and the {\em belief state} of an agent is represented by a logically closed set of (belief-representing) sentences. These sets are called {\em belief sets}. A change consists in adding or removing a specific sentence from a belief set to obtain a new belief set. The AGM model considers three kinds of belief change operators, namely {\em expansion, contraction} and {\em revision}.
An expansion occurs when new information is added to the set of the beliefs of an agent. The expansion of a belief set $\bs{K}$ by a sentence $\alpha$ (denoted by $\bs{K}+\alpha$) is the logical closure of $\bs{K}\cup \{\alpha\}$. A contraction occurs when information is removed from the set of beliefs of an agent. A revision occurs when new information is added to the set of the beliefs of an agent while retaining consistency if the new information is itself consistent. From the three operations, expansion is the only one that can be univocally defined. The other two operations are characterized by a set of postulates that determine the behaviour of each one of these functions, establishing conditions or constrains that they must satisfy.\\
Although the AGM model has acquired the status of standard model of belief change, several researchers (for an overview see \cite{FH11,FH18}) have pointed out its inadequateness in several contexts and proposed several extensions and generalizations to that framework. One of the criticisms to the AGM model that appears in the belief change literature is the total acceptance of the new information, which is characterized by the {\em success} postulate for revision.
``The AGM model always accepts the new information. This feature appears, in general, to be unrealistic, since rational agents, when confronted with information that contradicts previous beliefs, often reject it altogether or accept only parts of it" (\cite{FMT03}).
This may happen for various reasons. For example, the new information may lack on credibility or it may contradict previous highly entrenched beliefs. \\
Models in which the belief change operators considered do not satisfy the {\em success} postulate are designated by {\em non-prioritized belief change operators} (\cite{Han99}). The output of a non-prioritized revision may not contain the new belief that has motivated that revision. \\
Two level credibility-limited revision operators (two level CL revision operators for short) are non-prioritized revision operators that were proposed (independently) in \cite{Garapa22b} and \cite{Bon19}. When revising by means of a two level CL revision operator two levels of credibility and one level of incredibility are considered. When revising by a sentence at the highest level of credibility, the operator behaves as a standard revision. In this case the new information is incorporated in the agent's belief set. If the sentence is at the second level of credibility, then the outcome of the revision process coincides with a standard contraction by the negation of that sentence. In this case, the new information is not accepted but all the beliefs that are inconsistent with it are removed. The intuition underlying this behaviour is that, the belief is not credible enough to be incorporated in the agent's belief set, but creates some doubt in the agent's mind making her remove all the beliefs that are inconsistent with it.\\
In this paper, we propose a construction for {\em two level CL revision operators} based on Grove's systems of spheres and present an axiomatic characterization for these operators. The rest of the paper is organized as follows: In Section \href{set_back}{2} we introduce the notations and recall the main background concepts and results that will be needed throughout this article. In Section \href{sectTLCL1}{3} we present the two level CL revision operators and an axiomatic characterization for a class of these operators. In Section \href{set_SS}{4} we propose a construction for {\em two level CL revision operators} based on Grove's systems of spheres and present an axiomatic characterization for these operators. In Section \href{setrelated}{5}, we present a brief survey of related works. In Section \href{set_conc}{6}, we summarize the main contributions of the paper.

\section{Background}\label{set_back}

\subsection{Formal Preliminaries}

We will assume a propositional language $\LP$ that contains the usual truth functional connectives: $\lnot$ (negation), $\land$ (conjunction), $\lor$ (disjunction), $\to$ (implication) and $\leftrightarrow$ (equivalence). We will also use $\LP$ to denote the set of all formulas of the language. We shall make use of a consequence operation $Cn$ that takes sets of sentences to sets of sentences and which satisfies the standard Tarskian properties, namely {\em inclusion, monotony} and {\em iteration}. Furthermore, we will assume that $Cn$ satisfies {\em supraclassicality, compactness} and {\em deduction}. We will sometimes use $Cn(\alpha)$ for $Cn(\{\alpha\})$, $A \vdash \alpha$ for $\alpha \in Cn(A)$, $\vdash \alpha$ for $\alpha \in Cn(\emptyset)$, $A \not\vdash \alpha$ for $\alpha \not\in Cn(A)$, $\not\vdash \alpha$ for $\alpha \not\in Cn(\emptyset)$. The letters $\alpha, \beta, \ldots$ will be used to denote sentences of $\LP$.
$A, B, \ldots$ shall denote sets of sentences of $\LP$. $\bs{K}$ is reserved to represent a set of sentences that is closed under logical consequence (i.e. $\bs{K} = Cn(\bs{K})$) --- such a set is called a {\em belief set} or {\em theory}. Given a belief set $\bs{K}$ we will denote $Cn(\bs{K}\cup \{\alpha\})$ by $\bs{K}+\alpha$. We will use the symbol $\top$ to represent an arbitrary tautology and the symbol $\perp$ to represent an arbitrary contradiction. A possible world is a maximal consistent subset of $\LP$. The set of all possible worlds will be denoted by $\MLP$. Sets of possible worlds are called propositions. The set of possible worlds that contain $R\subseteq \LP$ is denoted by $\norm{R}$, i.e., $\norm{R}=\{M\in \MLP:R \subseteq M\}$. If $R$ is inconsistent, then $\norm{R}=\emptyset$. The elements of $R$ are designated by $R-$worlds. For any sentence $\alpha$, $\norm{\alpha}$ is an abbreviation of $\norm{Cn(\{\alpha\}})$ and its elements are designated by $\alpha$-worlds.

\subsection{AGM Revisions}\label{SubsecAGMcont}

The operation of revision of a belief set consists of the incorporation of new beliefs in that set. In a revision process, some previous beliefs may be retracted in order to obtain, as output, a consistent belief set. The following postulates, which were originally presented in \cite{Gar78,Gar82,Gar88}, are commonly known as {\em AGM postulates for revision}:\footnote{These postulates were previously presented in \cite{AGM85} but with slightly different formulations.}\\
\mypost{$\bsrev 1$} {${\bs{K}}\bsrev{\alpha} = Cn ({\bs{K}}\bsrev{\alpha})$ ({\em i.e.} ${\bs{K}}\bsrev{\alpha}$ is a belief set).}{Closure}
\mypost{$\bsrev 2$} {$\alpha \in {\bs{K}}\bsrev{\alpha}$.}{Success}
\mypost{$\bsrev 3$} {${\bs{K}}\bsrev{\alpha} \subseteq \bs{K}+\alpha$.}{Inclusion}
\mypost{$\bsrev 4$} {If $\neg\alpha \not\in \bs{K}$, then $\bs{K}+\alpha \subseteq {\bs{K}}\bsrev{\alpha}$.}{Vacuity}
\mypost{$\bsrev 5$} {If $\alpha$ is consistent, then $\bs{K} \bsrev \alpha$ is consistent.}{Consistency}
\mypost{$\bsrev 6$} {If $\vdash \alpha \leftrightarrow \beta$, then ${\bs{K}}\bsrev{\alpha} = {\bs{K}}\bsrev{\beta}$.}{Extensionality}
%
%
%
%
\mypost{$\bsrev 7$} {$\bs{K}\bsrev \alpha \cap \bs{K}\bsrev \beta \subseteq \bs{K}\bsrev (\alpha \vee \beta)$.}{Disjunctive overlap}
\mypost{$\bsrev 8$} {If $\neg \alpha \not \in \bs{K}\bsrev(\alpha \vee \beta)$, then $\bs{K}\bsrev (\alpha \vee \beta) \subseteq \bs{K}\bsrev \alpha$.}{Disjunctive inclusion}

\begin{definition}[\cite{AGM85}]\label{def_AGM_rev}
	An operator $\bsrev$ for a belief set $\bs{K}$ is a basic AGM revision if and only if it satisfies postulates {\bf $(\bsrev1)$} to {\bf $(\bsrev6)$}.
	It is an AGM revision if and only if it satisfies postulates {\bf $(\bsrev1)$} to  {\bf $(\bsrev8)$}.
\end{definition}

\subsection{AGM Contractions}\label{SubsAGMcon}

A contraction of a belief set occurs when some beliefs are removed from it (and no new beliefs are added). The following postulates, which were presented in \cite{AGM85} (following \cite{Gar78,Gar82}), are commonly known as {\em AGM postulates for contraction}:\\
\mypost{$\div1$} {${\bs{K}}\div{\alpha} = Cn ({\bs{K}}\div{\alpha})$ ({\em i.e.} ${\bs{K}}\div{\alpha}$ is a belief set).}{Closure}
\mypost{$\div2$} {${\bs{K}}\div{\alpha} \subseteq \bs{K}$.}{Inclusion}
\mypost{$\div3$} {If $\alpha \not\in \bs{K}$, then $\bs{K}\subseteq {\bs{K}}\div{\alpha}$.}{Vacuity}
\mypost{$\div4$} {If $\not\vdash\alpha$, then $\alpha \not\in {\bs{K}}\div{\alpha}$.}{Success}
\mypost{$\div5$} {$\bs{K} \subseteq ({\bs{K}}\div{\alpha}) + \alpha$.}{Recovery}
\mypost{$\div6$} {If $\vdash \alpha \leftrightarrow \beta$, then ${\bs{K}}\div{\alpha} = {\bs{K}}\div{\beta}$.}{Extensionality}
%
%
%
%
\mypost{$\div7$} {${\bs{K}}\div{\alpha} \cap {\bs{K}}\div{\beta} \subseteq {\bs{K}}\div{(\alpha\wedge\beta)}$.}{Conjunctive overlap}
\mypost{$\div8$} {${\bs{K}}\div{(\alpha\wedge\beta)} \subseteq {\bs{K}}\div{\alpha}$ whenever $\alpha \not\in {\bs{K}}\div{(\alpha\wedge\beta)}$.}{Conjunctive inclusion}

\begin{definition}[\cite{AGM85}]\label{def_AGM_con}
	An operator $\div$ for a belief set $\bs{K}$ is a basic AGM contraction if and only if it satisfies postulates {\bf $(\div1)$} to {\bf $(\div6)$}. It is an AGM contraction if and only if it satisfies postulates {\bf $(\div1)$} to {\bf $(\div8)$}.
\end{definition}

There are several contraction operators that are exactly characterized by the postulates {\bf $(\div1)$} to {\bf $(\div8)$}, namely the {\em (transitively relational) partial meet contractions} \cite{AGM85}, {\em safe contraction} \cite{AM85,RH14}, {\em system of spheres-based contraction} \cite{Gro88} and {\em epistemic entrenchment-based contraction} \cite{Gar88,GM88}.

The Levi and Harper identities\footnote{{\bf Harper identity:} \cite{Har76a} $\bs{K}\div\alpha=(\bs{K}\bsrev\neg\alpha)\cap \bs{K}$.\\{\bf Levi identity:} \cite{Lev77} $\bs{K}\bsrev\alpha=(\bs{K}\div\neg\alpha)+\alpha$.} make contraction and revision interchangeable. These identities allow us to define the revision and the contraction operators in terms of each other. The Levi (respectively Harper) identity enable the use of contraction (resp. revision) as primitive function and treat revision (resp. contraction) as defined in terms of contraction (resp. revision).

\subsection{Sphere-based Operations of Belief Change}

Grove (\cite{Gro88}), inspired by the semantics for counterfactuals (\cite{Lew73}) proposed  a structure called {\em system of spheres} to be used for defining revision functions. Figuratively, the distance between a possible world and the innermost sphere reflects its plausibility towards $\norm{\bs{K}}$. The closer a possible world is to $\norm{\bs{K}}$, the more plausible it is.

\begin{definition}[\cite{Gro88}]\label{def_sytem_spheresAGM}
	Let $\bs{K}$ be a belief set. A system of spheres, or spheres' system, centred on $\norm{\bs{K}}$ is a collection $\mathbb S$ of subsets of $\MLP$, i.e., $\mathbb S \subseteq \mathcal{P}(\MLP)$, that satisfies the following conditions:\\
	\mypostnoname{$\mathbb S 1$} {$\mathbb S$ is totally ordered with respect to set inclusion; that is, if $U, V\in \mathbb S$, then $U\subseteq V$ or $V \subseteq U$.}
	\mypostnoname{$\mathbb S 2$} {$\norm{\bs{K}} \in \mathbb S$, and if $U\in \mathbb S$, then $\norm{\bs{K}} \subseteq U$ ($\norm{\bs{K}}$ is the $\subseteq$-minimum of $\mathbb S)$.}
	\mypostnoname{$\mathbb S 3$} {$\MLP\in \mathbb S$ ($\MLP$ is the largest element of $\mathbb S$).}
	\mypostnoname{$\mathbb S 4$} {For every $\alpha \in \LP$, if there is any element in $\mathbb S$ intersecting $\norm{\alpha}$ then there is also a smallest element in $\mathbb S$ intersecting $\norm{\alpha}$.}
	
	The elements of $\mathbb S$ are called spheres. For any consistent sentence $\alpha\in \LP$, the smallest sphere in $\mathbb S$ intersecting $\norm{\alpha}$ is denoted by $\mathbb S_{\alpha}$.
\end{definition}

Given a system of spheres $\mathbb S$ centered on $\norm{\bs{K}}$ it is possible to define expansion, revision and contraction operators based on $\mathbb S$.

\begin{definition}[\cite{Gro88}]
	Let $\bs{K}$ be a belief set.

 (a)  An operation $+$ on $\bs{K}$ is a system of spheres-based expansion operator if and only if there exists a system of spheres $\mathbb S$ centered on $\norm{\bs{K}}$ such that for all $\alpha$ it holds that:
	\begin{equation*}
		\bs{K}+\alpha = \bigcap(\norm{\bs{K}}\cap\norm{\alpha}).
	\end{equation*}

(b) An operation $\div$ on $\bs{K}$ is a system of spheres-based contraction operator if and only if there exists a system of spheres $\mathbb S$ centered on $\norm{\bs{K}}$ such that for all $\alpha$ it holds that:
	\begin{equation*}
		\bs{K} \div\alpha = \left\{
		\begin{array}{ll}
			\bigcap((S_{\neg\alpha}\cap\norm{\neg \alpha}) \cup \norm{\bs{K}})   & \text{if }  \norm{\neg\alpha}\not=\emptyset\\
			\bs{K}& \textrm{otherwise}\\
		\end{array}
		\right.
	\end{equation*}

 (c) An operation $\bsrev$ on $\bs{K}$ is a system of spheres-based revision operator if and only if there exists a system of spheres $\mathbb S$ centered on $\norm{\bs{K}}$ such that for all $\alpha$ it holds that:
	\begin{equation*}
		\bs{K}\bsrev\alpha = \left\{
		\begin{array}{ll}
			\bigcap{(S_\alpha\cap\norm{\alpha})}   & \text{if }  \norm{\alpha}\not=\emptyset\\
			\LP& \textrm{otherwise}\\
		\end{array}
		\right.
	\end{equation*}
\end{definition}

It holds that sphere-based revision and contraction operators are characterized, by the (eight) AGM postulates for revision and contraction, respectively (\cite{Gro88}).

\section{Two Level Credibility-limited Revisions}\label{sectTLCL1}

The {\em two level CL revisions} are operators of non-prioritized revision. When revising a belief set by a sentence $\alpha$, we first need to analyse the degree of credibility of that sentence. When revising by a sentence that is considered to be at the highest level of credibility, the operator works as a standard revision operator. If it is considered to be at the second level of credibility, then that sentence is not incorporated in the revision process but its negation is removed from the original belief set.
When revising by a non-credible sentence, the operator leaves the original belief set unchanged.
The following definition formalizes this concept:

\begin{definition}[\cite{Garapa22b,Bon19}]
	Let $\bs{K}$ be a belief set, $\bsrev$ be a basic AGM revision operator on $\bs{K}$ and $C_{H}$ and $C_{L}$ be subsets of $\LP$. Then $\odot$ is a two level CL revision operator induced by $\bsrev$, $C_{H}$ and $C_{L}$ if and only if:
	\begin{equation*}
		\bs{K} \odot \alpha = \left\{
		\begin{array}{ll}
			\bs{K} \bsrev \alpha  & \text{if }  \alpha \in C_{H}\\
			(\bs{K} \bsrev \alpha)\cap \bs{K}  & \text{if }  \alpha \in C_{L}\\
			\bs{K}&  \text{if }  \alpha \not \in (C_{L}\cup C_{H})\\
		\end{array}
		\right.
	\end{equation*}
\end{definition}

In the previous definition $C_{H}\cup C_{L}$ represent the sentences that are considered to have some degree of credibility. $C_{H}$ and $C_{L}$ represent respectively  the set of sentences that are considered to be at the first (highest) and at the second level of credibility. Note that if $\alpha \in C_{L}$, then $\bs{K} \odot \alpha=(\bs{K} \bsrev \alpha)\cap \bs{K}$. According to the Harper identity
$(\bs{K} \bsrev \alpha)\cap \bs{K}$ coincides with the contraction of $\bs{K}$ by $\neg \alpha$.

This construction can be further specified by adding constraints to the structure of the set(s) of credible sentences. In \cite{HFCF01,GFR18a}, the following properties for a given set of credible sentences $C$  were proposed:

\noindent
{\bf Credibility of Logical Equivalents:} If $\vdash \alpha\leftrightarrow \beta$, then  $\alpha \in C$ if and only if $\beta \in C$.\footnote{In \cite{HFCF01} this property was designated by {\em closure under logical equivalence} and was formulated as follows: If $\vdash \alpha\leftrightarrow \beta$, and $\alpha \in C$, then $\beta \in C$.}    \\
{\bf Single Sentence Closure:} If $\alpha\in C$, then $Cn(\alpha)\subseteq C$.\\
{\bf Element Consistency:} If $\alpha \in C$, then $\alpha \not \vdash \perp$.\\
{\bf Credibility lower bounding:} If $\bs{K}$ is consistent, then $\bs{K}\subseteq C$.\\

Additionally, in \cite{Garapa22b} the following condition that relates a set of credible sentences $C$ with a revision function $\star$ was introduced. This condition, designated by {\em condition} (\ref{C-rev}), states that if a sentence $\alpha$ is not credible, then any possible outcome of revising the belief set $\bs{K}$ through $\star$ by a credible sentence contains $\neg \alpha$. The intuition underlying this property is that if $\alpha$ is not credible then its negation cannot be removed. Thus its negation should still be in the outcome of the revision by any credible sentence.
\vspace{-0.3cm}
\begin{equation*}
	\label{C-rev} \tag{\bf{C - $\star$}}
	\textrm{If }\alpha \not\in C \textrm{ and } \beta \in C\textrm{, then }  \neg \alpha \in \bs{K} \star \beta.
\end{equation*}

\subsection{Two level credibility-limited revision postulates }

We now recall from \cite{Garapa22b} some of the postulates proposed to express properties of the two level CL revision operators.
The first postulate was originally proposed in \cite{Mak97b}, the second in \cite{KM91},  the following three in \cite{HFCF01} and the remaining ones in \cite{Garapa22b}.

\mypostnoname{Consistency Preservation} {If $\bs{K}$ is consistent, then $\bs{K}\odot\alpha$ is consistent.}{}
\mypostnoname{Confirmation} {If $\alpha \in \bs{K}$, then $\bs{K}\odot \alpha =\bs{K}$.}{}
\mypostnoname{Strict Improvement} {If $\alpha \in \bs{K}\odot \alpha$ and $\vdash \alpha \rightarrow \beta$, then $\beta \in \bs{K}\odot \beta$.}{}
\mypostnoname{Regularity} {If $\beta \in \bs{K}\odot \alpha$, then $\beta \in \bs{K}\odot \beta$.}{}
\mypostnoname{Disjunctive Distribution} {If $\alpha\vee \beta\in \bs{K}\odot (\alpha \vee \beta)$, then $\alpha \in \bs{K}\odot \alpha$ or $\beta \in \bs{K}\odot \beta$.}{}
\mypostnoname{N-Recovery} {$\bs{K}\subseteq \bs{K}\odot\alpha+\neg \alpha$.}
\mypostnoname{N-Relative success} {If $\neg\alpha \in \bs{K}\odot\alpha$, then $\bs{K}\odot\alpha=\bs{K}$.}{}
\mypostnoname{N-Persistence} {If $\neg \beta \in \bs{K}\odot\beta$, then $\neg \beta \in  \bs{K}\odot\alpha$.}
\mypostnoname{N-Success Propagation} {If $\neg \alpha \in \bs{K}\odot \alpha$ and $\vdash \beta \rightarrow \alpha$, then $\neg \beta \in \bs{K}\odot\beta$.}{}
\mypostnoname{Weak Relative Success} {$\alpha \in \bs{K}\odot\alpha$ or $\bs{K}\odot \alpha\subseteq \bs{K}$.}{}
\mypostnoname{Weak Vacuity} {If $\neg\alpha \not\in \bs{K}$, then $\bs{K} \subseteq {\bs{K}}\odot{\alpha}$.}{}
\mypostnoname{Weak Disjunctive Inclusion} {If $\neg \alpha \not \in \bs{K}\odot(\alpha \vee \beta)$, then $\bs{K}\odot (\alpha \vee \beta)+(\alpha \vee \beta)\subseteq \bs{K}\odot \alpha+\alpha$.}{}
\mypostnoname{Containment} {If $\bs{K}$ is consistent, then $\bs{K}\cap ((\bs{K}\odot\alpha)+\alpha)\subseteq \bs{K}\odot\alpha$.}{}

The following observations relate some of the postulates presented above.

\begin{observation}[\cite{Garapa22b}]\label{Nrelat_suc}\label{k=koa}
	Let $\bs{K}$ be a consistent and logically closed set and $\odot$ be an operator on $\bs{K}$. \\
	(a) If $\odot$ satisfies closure, consistency preservation, weak relative success and N-Recovery, then it satisfies N-Relative success.\\ (b) If $\odot$ satisfies  weak vacuity and inclusion, then it satisfies confirmation.
\end{observation}


\begin{observation}\label{obs_containment1}
	Let $\bs{K}$ be a consistent and logically closed set and $\odot$ be an operator on $\bs{K}$. \\
(a) If $\odot$ satisfies consistency preservation, closure, vacuity, inclusion, strict improvement, disjunctive inclusion, disjunctive overlap and N-recovery, then it satisfies regularity.\\
%
 (b) If $\odot$ satisfies consistency preservation, closure, vacuity, weak relative success and disjunctive inclusion, then it satisfies disjunctive distribution.\\
%
(c) If $\odot$ satisfies N-recovery and closure, then it satisfies containment.
\end{observation}

In the following theorem we recall from \cite{Garapa22b} an axiomatic characterization for a two level CL revision operator induced by an AGM revision and sets $C_{H}$ and $C_{L}$ satisfying some given properties.\footnote{Actually, the containment postulate was also included in the list of postulates of the representation theorem presented in \cite{Garapa22b}, however as Observation \ref{obs_containment1} illustrates, containment follows from closure and N-recovery.}

\begin{observation}[\cite{Garapa22b}]\label{Rep_Tcl_AGM}
	Let $\bs{K}$ be a consistent and logically closed set and $\odot$ be an operator on $\bs{K}$. Then the following conditions are equivalent:

1. $\odot$ satisfies weak relative success, closure, inclusion, consistency preservation, weak vacuity, extensionality, strict improvement, N-persistence, N-recovery, disjunctive overlap and weak disjunctive inclusion.

2. $\odot$ is a two level CL revision operator induced by an AGM revision operator $\star$ for $\bs{K}$ and sets $C_{H}, C_{L}\subseteq \LP$ such that:
		$C_{L}$ satisfy credibility of logical equivalents and element consistency, $C_{H}\cap C_{L}=\emptyset$, $C_{H}$ satisfies element consistency, credibility lower bounding and single sentence closure and condition {\bf ($C_{H}\cup C_{L}$ - $\star$)} holds.
\end{observation}

\section{System of Spheres-based Two Level Credibility-limited Revisions}\label{set_SS}

In this section we present the definition of a system of spheres-based two level CL revision operator. We start by presenting the notion of two level system of spheres, centred on $\norm{\bs{K}}$.

\begin{definition}\label{def_SiS}
	Let $\bs{K}$ be a belief set. A two level system of spheres centred on $\norm{\bs{K}}$ is a pair $(\mathbb S_i,\mathbb S)$ whose elements are subsets of $\MLP$, i.e., $\mathbb S \subseteq \mathcal{P}(\MLP)$ and $\mathbb S_i \subseteq \mathcal{P}(\MLP)$, such that:

		(a) 	$\mathbb S$ and $\mathbb S_i$ satisfy conditions $(\mathbb S 1)$, $(\mathbb S 2)$ and $(\mathbb S 4)$ of Definition \ref{def_sytem_spheresAGM};

		(b) $\mathbb {S}_i \subseteq \mathbb S$;

		(c) If $X\in \mathbb S_i$, then $X\subseteq Y$ for all $Y\in \mathbb S\setminus \mathbb S_i$.
\end{definition}

\begin{figure}[ht]
	\centering
	\includegraphics[scale=0.55]{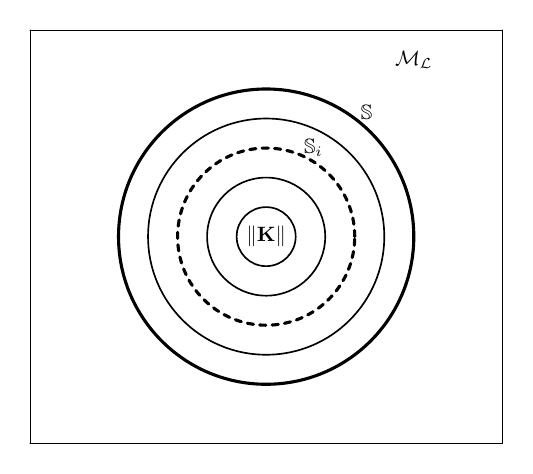}
	\caption{Schematic representation of a two level system of spheres $(\mathbb S_i,\mathbb S)$, centred on $\norm{\bs{K}}$. The dashed circle establishes the boundary between the spheres of $\mathbb S_i$ and those of $\mathbb S \setminus\mathbb S_i$. Worlds outside the thickest line are not elements of any sphere of $\mathbb S$.}
	\label{fig1}
\end{figure}

Intuitively, a two level system of spheres $(\mathbb S_i,\mathbb S)$, centered on $\norm{\bs{K}}$ is a system composed by two systems of spheres $\mathbb S_i$ and $\mathbb S$, both centered on $\norm{\bs{K}}$, where $\mathbb {S}_i \subseteq \mathbb S$ and in which the condition $(\mathbb S 3)$ of Definition \ref{def_sytem_spheresAGM} is relaxed for $\mathbb S_i$ and $\mathbb S$, allowing the existence of possible worlds outside the union of all spheres of $\mathbb S_i$ and of $\mathbb S$.\footnote{Condition $(\mathbb S 3)$ of Definition \ref{def_sytem_spheresAGM} was also relaxed in \cite{HFCF01} when constructing a (modified) system of spheres for credibility-limited revision operators.} Conditions (b) and (c) impose that the spheres of $\mathbb S_i$ are the innermost ones (see Figure \ref{fig1}).

The following observation is a direct consequence of condition (c). It states that all spheres contained in a given sphere of $\mathbb S_i$ belong to $\mathbb S_i$.

\begin{observation}\label{obsC}
	If $\mathbb S_i$ and $ \mathbb S$ satisfy condition (c) of Definition \ref{def_SiS}, then it holds that:\\
	If $X\in \mathbb S$ and $Y\in 	\mathbb S_i$ are such that $X\subseteq Y$, then $X\in \mathbb S_i$.
\end{observation}

In a system of spheres centered on $\|\bs{K}\|$, the worlds considered most plausible are those that lie in the innermost sphere (i.e. in $\|\bs{K}\|$), and the closer a possible world is to the center, the more plausible it is considered to be. Similarly, the worlds lying in the spheres of $\mathbb S_i$ have a higher degree of plausibility than those in the spheres of $\mathbb S\setminus\mathbb S_i$.
Intuitively, a two level system of spheres $(\mathbb S_i,\mathbb S)$, centered on $\norm{\bs{K}}$ defines three clusters. The first cluster is formed by the worlds in the spheres of $\mathbb S_i$. These worlds are the ones to which a higher degree of plausibility is assigned (relatively to those outside the spheres of $\mathbb S_i$). The second cluster is formed by the worlds in the spheres of $\mathbb S\setminus \mathbb S_i$, which are assigned some (lower) degree of plausibility. Finally, the third cluster is formed by the worlds outside the spheres of $\mathbb S$, which are considered to be not plausible.

\begin{figure}[ht]
	\centering
	\includegraphics[scale=0.5]{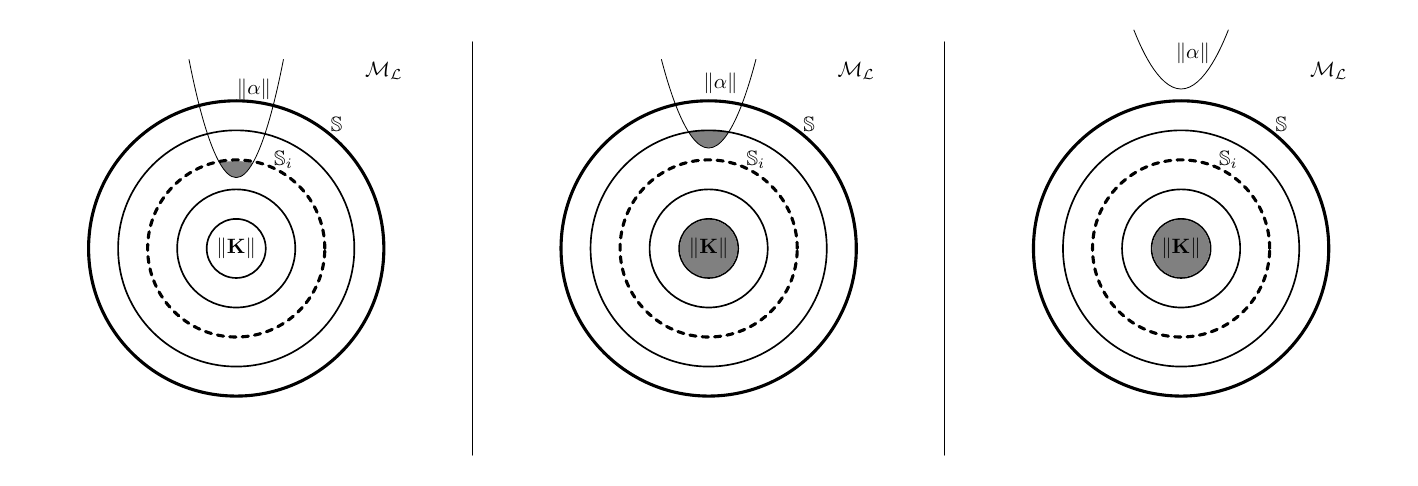}
	\caption{Schematic representation of the worlds of the outcome of the system of spheres-based two level CL revision operator induced by a two level system of spheres $(\mathbb S_i,\mathbb S)$ centred on $\norm{\bs{K}}$ by a sentence $\alpha$. In the first case, it holds that $S_\alpha\in \mathbb S_i$ and in the second that $S_\alpha\in \mathbb S \setminus \mathbb S_i$. In the third case, all the  $\alpha$-worlds are outside the spheres of $\mathbb S$.}
	\label{fig2}
\end{figure}

We are now in conditions to present the definition of a system of spheres-based two level CL revision operator. The outcome of the revision by means of a system of spheres-based two level CL revision operator of a belief set $\bs{K}$ by a sentence $\alpha$ (see Figure \ref{fig2}) is:\\
- the intersection of the most plausible $\alpha$-worlds, if these are $\alpha$-worlds in the cluster of the most plausible worlds.\footnote{Note that being $X$ a set of possible worlds $\bigcap X$ is a belief set.}\\
- the intersection of all the worlds contained in the union of the set of $\bs{K}$-worlds with the set of the most plausible $\alpha$-worlds, if the $\alpha$-worlds are considered to be plausible, but are not in the cluster of the most plausible ones. \\
- $\bs{K}$  if the $\alpha$-worlds are not plausible, i.e, in this case the belief set remains unchanged.

\begin{definition}
	Let $\bs{K}$ be a belief set and $(\mathbb S_i,\mathbb S)$ be a two level system of spheres centered on $\norm{\bs{K}}$. The system of spheres-based two level CL revision operator induced by $(\mathbb S_i,\mathbb S)$ is the operator $\odot_{(\mathbb S_i,\mathbb S)}$ such that, for all $\alpha$:
	\begin{equation*}
		\bs{K} \odot_{(\mathbb S_i,\mathbb S)} \alpha = \left\{
		\begin{array}{ll}
			\bigcap (S_\alpha\cap \norm{\alpha})  & \text{if }  S_\alpha \in \mathbb S_i\\
			\bigcap (\norm{\bs{K}}\cup (S_\alpha\cap \norm{\alpha}))  & \text{if }   S_\alpha \in \mathbb S\setminus \mathbb S_i\\
			\bs{K}&  \text{if } X\cap \norm{\alpha}=\emptyset,
			                     \text{for all } X\in  \mathbb S   \\
		\end{array}
		\right.
	\end{equation*}
	An operator $\odot$ on $\bs{K}$ is a system of spheres-based two level CL revision operator if and only if there exists a two levels system of spheres $(\mathbb S_i,\mathbb S)$ centred on $\norm{\bs{K}}$ such that $\bs{K} \odot\alpha=\bs{K} \odot_{(\mathbb S_i,\mathbb S)} \alpha$ holds for all $\alpha$.
\end{definition}

\subsection{Representation theorems}

We now present a representation theorem for system of spheres-based two level CL revision operators. It also relates these operators with the two level CL revision operators induced by AGM revision operators and sets $C_{H}, C_{L}\subseteq \LP$ satisfying some given properties. Considering the axiomatic characterization for the latter, presented in Observation \ref{Rep_Tcl_AGM}, we note that we only need to ensure that the Condition {\bf ($C_{H}$ - $\star$)} holds, to guarantee that the class of these operators coincides with the class of system of spheres-based two level CL revision operators.

\begin{theorem}\label{teo_main}
	Let $\bs{K}$ be a consistent and logically closed set and $\odot$ be an operator on $\bs{K}$. Then the following conditions are equivalent:

	1. $\odot$ satisfies weak relative success, closure, inclusion, consistency preservation, vacuity, extensionality, strict improvement, N-persistence, N-recovery, disjunctive overlap and disjunctive inclusion.

		2. $\odot$ is a system of spheres-based two level CL revision operator.

		3. $\odot$ is a two level CL revision operator induced by an AGM revision operator $\star$ for $\bs{K}$ and sets $C_{H}, C_{L}\subseteq \LP$ such that:
		$C_{L}$ satisfy credibility of logical equivalents and element consistency, $C_{H}\cap C_{L}=\emptyset$, $C_{H}$ satisfies element consistency, credibility lower bounding and single sentence closure and conditions {\bf ($C_{H}\cup C_{L}$ - $\star$)} and {\bf ($C_{H}$ - $\star$)} hold.
\end{theorem}

\section{Related Works}\label{setrelated}

In this section we will mention other approaches related with the present paper. \\
- In \cite{Garapa22b}, the two level CL revision operators were defined in terms of a basic AGM revision operator and sets $C_H$ and $C_L$ of credible sentences. Several properties have been proposed for these sets. Postulates to characterize two level CL revision operators were proposed. Results exposing the relation between the postulates and the properties of $C_H$ and $C_L$ were presented. Axiomatic characterizations for several classes of two level CL revision operators were presented (namely for two level CL revision operators induced by basic AGM revisions and by AGM revisions in which the associated sets of credible sentences satisfy certain properties).\\
- In \cite{Bon19}, the operators of two CL revision were introduced in terms of basic AGM belief revisions operators (in that paper these operators are designated by {\em Filtered belief revision}). The possibility that an item of information could still be ``taken'' seriously, even if it is not accepted as being fully credible (this type of information is there called {\em allowable}) was discussed. A syntactic analysis of filtered belief revision was provided.\\
- In \cite{Bon22}, the works presented in \cite{Bon19} and \cite{Garapa22b} were extended by introducing the notion of partial belief revision structure, providing a characterization of filtered belief revision in terms of properties of these structures.
	There it is considered the notion of rationalizability of a choice structure in terms of a plausibility order and established a correspondence between rationalizability and AGM consistency in terms of the eight AGM postulates for revision. An interpretation of credibility, allowability and rejection of information in terms of the degree of implausibility of the information was provided.\\
- In \cite{HFCF01} credibility-limited revision operators were presented. When revising a belief set by a sentence by means of a credibility-limited revision operator, we need first to analyse whether that sentence is credible or not. When revising by a credible sentence, the operator works as a basic AGM revision operator, otherwise it leaves the original belief set unchanged. Two level credibility-limited revisions operators can be seen as a generalization of credibility-limited revision operators. In fact, in the case that $C_L=\emptyset$ both types of operators coincide.
	%
	%
	In \cite{HFCF01} several properties were prosed for $C$ (the set of credible sentences) and this model was developed in terms of possible world models. Representations theorems for different classes of   Credibility-limited revisions operators were presented. The extension of credibility-limited revision operators to the belief bases setting was studied in \cite{FMT03,GFR18a,GFR23,GFR20}.

\section{Conclusion}\label{set_conc}

The model of credibility-limited revision (\cite{HFCF01}) is essentially a generalization of the AGM framework (\cite{AGM85}) of belief revision, which addresses one of the main shortcomings pointed out to that framework, namely the fact that it assumes that any new information has priority over the original beliefs. In the model of credibility-limited revisions two classes of sentences are considered. Some sentences ---the so-called {\em credible sentences}--- are accepted in the process of revision by them, while the remaining sentences are such that the process of revising by them has no effect at all in the original belief set.

In its turn, the model of two level CL revision (\cite{Bon19,Garapa22b}) generalizes credibility-limited revision by considering an additional class of sentences. A sentence of this class is such that, although a revision by it does not lead to its acceptance, it causes the removal of its negation from the original belief set.

The present paper offers a semantic approach to the two level CL revision operators. More precisely, it introduces a class of two-level CL revision operators whose definition is based on a structure called two level system of spheres, which generalizes the well-known systems of spheres proposed by Grove (\cite{Gro88}). This semantic definition provides some additional insight on the intuition that underlays the notion of two-level CL revisions.

\paragraph{Acknowledgements}

This paper was partially supported by FCT-Funda\c{c}\~{a}o para a Ci\^{e}ncia e a Tecnologia, Portugal through project PTDC/CCI-COM/4464/2020.
M.G. and M.R. were partially supported by the Centro de Investiga\c{c}\~{a}o em Matem\'{a}tica e Aplica\c{c}\~{o}es (CIMA), through the grant UIDB/04674/2020 of FCT. E.F. was partially supported by FCT through project UIDB/04516/2020 (NOVA LINCS).


\nocite{*}
\bibliographystyle{eptcs}

\section{Appendix}
In this appendix we provide a sketch proof for the main result presented in this paper.
\begin{proof}[Proof sketch  of Theorem \ref{teo_main}:]\hfill
	
 {\bf(2) to (1)}: \\
Let $\odot$ be a system of spheres-based two level credibility limited revision operator induced by a two levels system of spheres $(\mathbb S_i,\mathbb S)$. We need to prove that $\odot$ satisfies all the postulates present in statement (1) .

{\bf(1) to (2)}: \\
Assume 	that $\odot$ satisfies all the postulates listed in statement (1) and 	
consider the following constructions for $\mathbb S$ and $\mathbb S_i$:

$S\in \mathbb S_i$ iff:\\
	(a) $S=\norm{\bs{K}}$;\\
	(b) $\emptyset\not=S\subseteq \{w:w\in \norm{\bs{K}\odot \alpha}\text{, for some } \alpha \text{ such that} \norm{\bs{K}\odot \alpha}\subseteq \norm{\alpha}\}$ and $\norm{\bs{K}\odot \alpha}\subseteq S$ for all $\alpha$ such that $S\cap \norm{\alpha}\not=\emptyset$.\\

$S\in \mathbb S$ iff:      \\
	(a) $S=\norm{\bs{K}}$;\\
	(b) $\emptyset\not=S\subseteq \{w:w\in \norm{\bs{K}\odot \alpha}\text{, for some } \alpha \text{ such that} \norm{\bs{K}\odot \alpha}\cap \norm{\alpha}\not=\emptyset\}$, $\norm{\bs{K}\odot \alpha}\subseteq S$ for all $\alpha$ such that $S\cap \norm{\alpha}\not=\emptyset$ and if $S\cap \norm{\alpha}=\emptyset$ and $S\not \in \mathbb S_i$, then $\norm{\bs{K}\odot \alpha}\cap S=\norm{\bs{K}}$.\\


We need to show that:\\
	1.  $(\mathbb S_i,\mathbb S)$ is a two level system of spheres centred on $\norm{\bs{K}}$. To do so, it is necessary to prove that:
	\begin{enumerate}	
		\item[i.]  $\mathbb S$ and $\mathbb S_i$ satisfy conditions $(\mathbb S 1)$, $(\mathbb S 2)$ and $(\mathbb S 4)$, of Definition \ref{def_sytem_spheresAGM};
		\item[ii.] $\mathbb {S}_i \subseteq \mathbb S$;
		\item[iii.] If $X\in \mathbb S_i$, then $X\subseteq Y$ for all $Y\in \mathbb S\setminus \mathbb S_i$.
	\end{enumerate}	
	2. If $\norm{\alpha}=\emptyset$, then $\bs{K} \odot \alpha=\bs{K}$;\\
	3. For $\alpha$ such that $\bs{K}\odot\alpha\not\vdash \neg \alpha$ and $S(\alpha)=\bigcup\{\norm{\bs{K}\odot \delta}:\norm{\alpha}\subseteq \norm{\delta}\}$, it holds that:
	\begin{enumerate}
		\item[i.] $S(\alpha)\in \mathbb S$
		\item[ii.] $S(\alpha)=S_\alpha$ (i.e. $S(\alpha)$ is the minimal sphere that intersects with $\norm{\alpha}$).
		\item[iii.]
		\begin{equation*}
			\bs{K} \odot \alpha = \left\{
			\begin{array}{ll}
				\bigcap (S_\alpha\cap \norm{\alpha})  & \text{if }  S_\alpha \in \mathbb S_i\\
				\bigcap (\norm{\bs{K}}\cup (S_\alpha\cap \norm{\alpha}))  & \text{if }   S_\alpha \in \mathbb S\setminus \mathbb S_i\\
				\bs{K}&  \text{if } X\cap \norm{\alpha}=\emptyset,
				\text{for all } X\in  \mathbb S   \\
			\end{array}
			\right.,
		\end{equation*}
		where $S_\alpha=S(\alpha)$.
	\end{enumerate}
	%

{\bf(1) to (3)}: \\
Let $\odot$ be an operator satisfying the postulates listed in statement (1).
Let $\star$ be the operation such that:
\begin{itemize}
	\item[i.] If $\neg \alpha \not \in \bs{K}\odot\alpha$, then $\bs{K}\star\alpha=\bs{K}\odot\alpha+\alpha$;
	\item[ii.] If  $\neg \alpha \in \bs{K}\odot\alpha$, then $\bs{K}\star\alpha=Cn(\alpha)$.
\end{itemize}
Furthermore let $C_{H}=\{\alpha: \alpha \in \bs{K}\odot\alpha\}$ and $C_{L}=\{\alpha: \neg \alpha \not \in \bs{K}\odot\alpha\}\setminus C_{H}$.\\
These are the same construction that were used in the corresponding part of Observation \ref{Rep_Tcl_AGM}. Then, regarding this proof, it remains only to show that condition {\bf ($C_{H}$ - $\star$)} holds.

{\bf(3) to (1)}: \\
By Observation \ref{Rep_Tcl_AGM} it
only remains to prove that $\odot$ satisfies vacuity and disjunctive inclusion.
\end{proof}

\end{document}